\documentclass[letterpaper,peerreviewca,11pt,draftcls,oneside]{IEEEtran}
\usepackage[mathcal]{euscript}
\usepackage{amssymb,amsmath,amsthm}
\usepackage{times}
\usepackage{graphicx}
\usepackage{epstopdf}
\usepackage{graphicx}
\usepackage{amsfonts}
\usepackage{color}
\usepackage[T1]{fontenc}
\usepackage{mathtools}

\title{A Multi-Service Oriented Multiple-Access Scheme For M2M Support in Future
LTE}
\author{Nassar Ksairi, Stefano Tomasin and M\'erouane Debbah\\
Mathematical and Algorithmic Sciences Lab,
France Research Center,\\
Huawei Technologies Co. Ltd., Boulogne-Billancourt, France.\\
Emails:
\{nassar.ksairi, stefano.tomasin, merouane.debbah\}@huawei.com}
\date{July 2016}

\begin{document}
\maketitle
\begin{abstract}
We propose a novel multiple-access technique to overcome the
shortcomings of the current proposals for the future releases of Long-Term
Evolution (LTE). We provide a unified radio access system that efficiently and
flexibly integrates both traditional cellular services and machine-to-machine
(M2M) connections arising from Internet-of-Things (IoT) applications.
The proposed solution, referred to as multi-service oriented multiple access
(MOMA), is based on a) establishing separate classes of users using relevant
criteria that go beyond the simple handheld-IoT device split,
b) service-dependent  hierarchical spreading of the data signals and c) a mix of
multiuser and single-user detection schemes at the receiver. Signal spreading in
MOMA allows to handle densely connected devices with different
quality-of-service (QoS) profiles and at the same time its flexible receiver
structure allows to allocate the receiver computational resources to the
connections that need it most. This yields a scalable and efficient use of the
available radio resources and a better service integration. While providing
significant advantages for key future communications scenarios, MOMA can be
incorporated into LTE with a limited impact on the protocol structure and the
signaling overhead.
\end{abstract}

\section*{Introduction}

The provisioning of IoT services is now widely seen in the telecommunications
sector as one of the major features in the evolution of cellular systems.
Indeed, having a unified cellular system capable of handling
both IoT machines and handheld mobile devices would be greatly advantageous for
both operators and users. Towards this goal,
the design of an integrated radio access solution is a challenging problem.
Major issues are the large number of IoT devices required to be simultaneously
served and the difference between their traffic patterns/QoS requirements and
those of mobile broadband services~\cite{smart_city}. Another issue is that
IoT applications do not all have the same QoS and traffic
characteristics~\cite{ngmn}, with the implication that future M2M-related system
optimization should have inherent flexible support for several types of IoT
services.

To address some of these challenges, the Third Generation Partnership Project
(3GPP) has started to add M2M-type communications into the radio access
subsystem of LTE starting from Release 12~\cite{rel_12} by
introducing a new user equipment (UE) category, namely Category~0 (Cat.~0).
Cat.~0 devices are characterized by their low cost due to their by-design lack
of support for high peak rates and multiple antennas.
For Release 13 of the LTE standard, 3GPP is working on providing further cost
reductions for M2M communications~\cite{rel_13}. The proposals that emerged
within this work are referred to as {\it clean-slate narrow-band
cellular IoT} (NB CIoT)~\cite{ciot} and most of them advocate orthogonal
physical-layer transmission schemes that are a mixture of frequency division
multiple access (FDMA) and time division multiple access (TDMA).
This is the case, for example, of LTE for Machine-Type Communications (LTE-M)
and Narrow-band LTE-M (NB LTE-M).
Each of these two proposals introduces a new UE category, the so-called Cat.~1.4
MHz for LTE-M and Cat.~200 kHz for NB LTE-M~\cite{lte_m}. As their respective
names indicate, these new UE categories restrict the device transceiver
bandwidth to 1.4 MHz and 200 kHz, respectively. Other proposals focused on
upgrading the LTE random access procedure for better support of massive M2M
transmissions~\cite{lte_ra} by overcoming the so-called Physical Random-Access
Channel (PRACH) overloading issue.

\section*{Limitations of M2M Proposals for Next LTE Releases}

None of the existing M2M-related proposals for LTE is able to meet the following
crucial requirements all at once.
\begin{enumerate}
 \item {\bf Multi-class users/services:}
While most of the existing proposals treat IoT devices as a single class
of users, not enough attention has been paid to the different QoS and traffic
profiles within this class. Indeed, IoT services \emph{will not be limited to
data collection from simple sensors and will not only emit small 
data packets}~\cite{rel_13,ngmn}. One example is mobile video surveillance which
is expected to use medium to high-end devices that do not have battery life
constraints~\cite{ngmn}. Moreover, there should be a distinction within
the services running on handheld devices between mobile broadband applications
and the applications that have traffic characteristics and data rate
requirements resembling those typical of IoT services. In the latter group we
have  for example the messages generated by social networking and chatting
applications. Significant gains in resource utilization efficiency are expected
from a multiple-access scheme that treats the services with similar QoS
profiles, whether running on handhelds or IoT machines, as belonging to the same
user class.
 \item {\bf Dense IoT deployment:}
The existing M2M proposals for LTE allow to enhance its IoT capabilities but
their reliance on FDMA limits the number of simultaneous M2M connections to 
the (typically moderate) number of frequency sub-channels
they reserve for IoT communications.
There is a need to support a much larger number of simultaneously connected IoT
devices (and possibly mobile services with low QoS requirements). This goal
should be met without sacrificing the QoS of mobile broadband services.
 \item {\bf Flexibility in resource assignment:}
In existing M2M proposals, there is no way to dynamically adjust the
respective proportions of resources assigned to broadband services and IoT
devices. This could lead to wasting resources or to denials of
service depending on the current traffic demands.
Moreover, these proposals have limited flexibility in resource
allocation within the M2M frequency band which only comes from varying the
number of sub-channels occupied by each device from within a limited number of
possible values (6 in LTE-M). Any new multiple-access scheme should be more
flexible in assigning resources to different classes and to users within each
class.
 \item {\bf Efficiency in resource utilization:}
Assigning orthogonal frequency sub-channels to devices with low QoS requirements
is not the most efficient way to use the available radio resources. First,
this will bound the maximum number of simultaneously connected devices by the
number of available sub-channels. Second, it is known that the boundary of the
multiple-access channel (MAC) capacity region is not achieved with orthogonal
transmission schemes. Third, achieving robustness against timing and carrier
frequency offsets in time and/or frequency division orthogonal schemes requires
the use of guard intervals and/or bands around each sub-channel, further
reducing their resource utilization efficiency.
\end{enumerate}
In the sequel, we show that MOMA can overcome these limitations while being
compatible with low-cost devices having simple transceivers and long battery
life requirements. Indeed, both the (narrow) bandwidth values that were
originally proposed for LTE-M (1.4 MHz) and NB LTE-M (200 kHz) as a way to
reduce transceiver complexity are supported in MOMA.

\section*{Multi-Service Oriented Multiple Access}

MOMA is a multiple-access scheme conceived for scenarios where users are grouped
into different classes. It reveals all its potentials when the BS is equipped
with a large number $M$ of antennas. In this article we assume that classes
are defined based on users' QoS requirements profiles. For example, we
define $L\geq2$ classes of users as follows.
\begin{itemize}
\item {\bf One maximum data rate (HD) class of users:} Here, HD stands for
{\em high data rate}. For the HD class the objective is to obtain
\emph{a data rate as high as possible for $K^{\mathrm{HD}}$
simultaneous transmissions}. Typically these users are associated with
data-hungry applications on handheld devices such as video conferencing and
media streaming. 
\item $\mathbf{L-1}$ {\bf Constant low-to-moderate data rate (LMD) classes of
users:} Here LMD stands for {\em low-to-moderate data rate}. The $l$-th class,
with $l\in\{1,\ldots,L-1\}$, includes users requesting services with a
relatively low or moderate data rate $r_l^{\mathrm{LMD}}$. In the sequel
we assume that LMD classes are ordered from $l=1$ to $l=L-1$ with increasing
target data rates.
Services with low target rates could originate from applications running either
on handheld devices (such as social messaging) or on machines (such as M2M
light-duty data collection from smart meters and remote sensors).
The same applies to services with higher target data rates such as
moderate-quality live streaming from handheld devices and M2M heavy-duty data
collection from mobile video surveillance machines. Class
$l\in\{1,\ldots,L-1\}$ aims at accommodating the maximum number 
\emph{of simultaneous transmissions $K_l^{\mathrm{LMD}}$ at the 
granted data rate $r_l^{\mathrm{LMD}}$}.
\end{itemize}

\begin{figure}
 \centering
 \includegraphics[width=0.9\hsize]{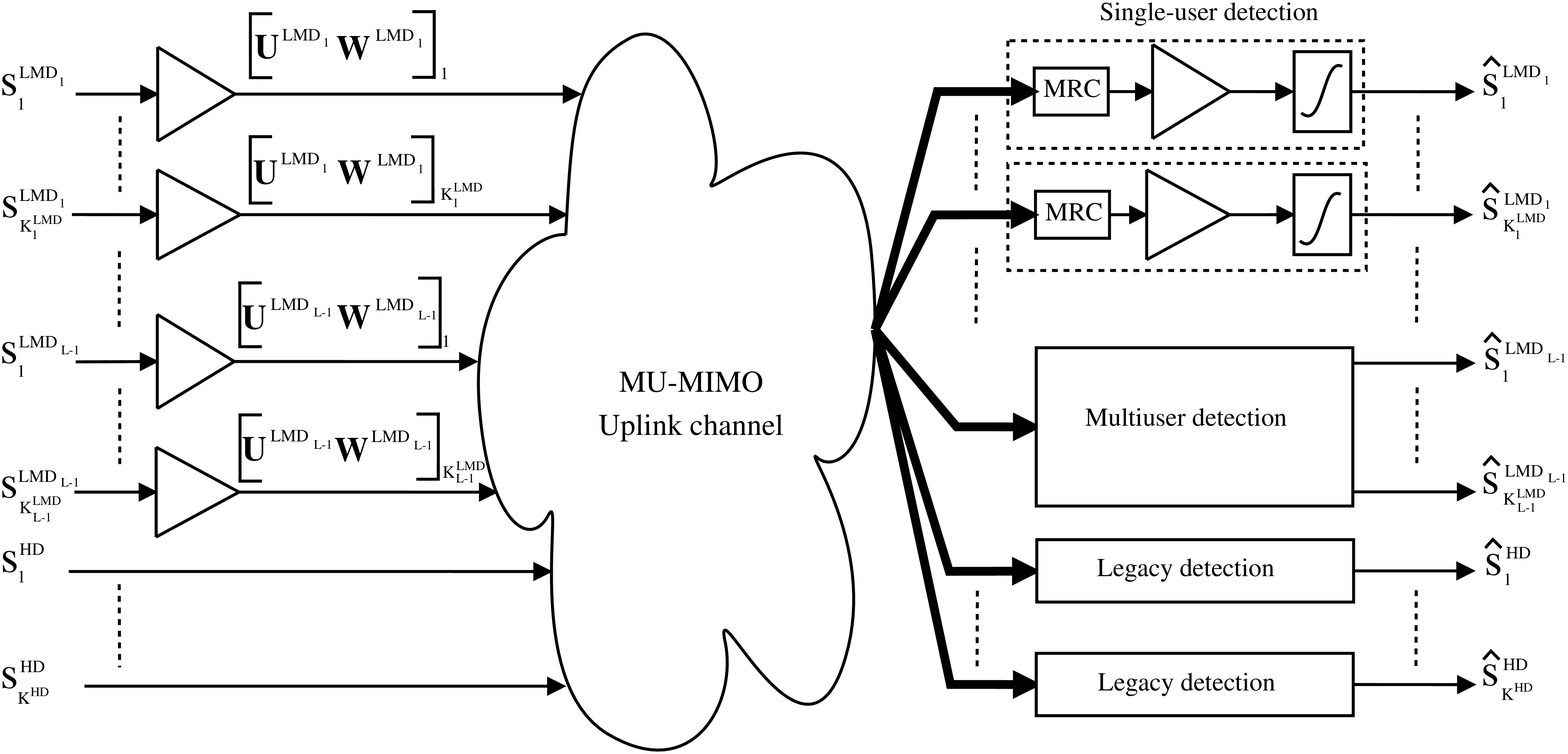}
 \caption{MOMA transceivers for $L$ classes of users.
$[\mathbf{M}]_j$ designates the $j$-th column of matrix $\mathbf{M}$.}
 \label{fig:moma_principle}
\end{figure}

Since what matters for HD users is maximizing their respective throughput,
proper scheduling techniques will typically limit the number of simultaneous
HD transmissions, exactly as in current wireless-communications standards.
It is thus reasonable to assume that $K^{\mathrm{HD}}$ is small and that
multiuser multiple-input multiple-output (MU-MIMO) techniques implemented on top
of orthogonal frequency division multiple access (OFDMA) in the downlink and
single-carrier frequency division multiple access (SC-FDMA) in the uplink can be
used for HD/HD signal separation. We also propose the use of these frequency
domain transmission schemes for the separation between the HD and the
other user classes. This choice allows to maintain full compatibility with
the LTE standard. As for the $L-1$ LMD classes, due to both their specific data
rate requirements and the objective of massive M2M deployment, we propose to
overload radio resources. Note that this overloading can be achieved,
for instance, by MU-MIMO techniques operating also in the code
domain. The way we propose to access the code domain is dubbed
{\em service dependent hierarchical spreading} and can be thought of as a
{\em layered} or {\em hierarchical} spreading with a class dependent overloading
factor. The MOMA uplink transmission scheme is illustrated in the left-hand part
of Fig.~\ref{fig:moma_principle}.

\subsection*{MOMA Features}
\begin{enumerate}
 \item MOMA is based on service dependent hierarchical spreading. This new
transmission scheme has the advantages of efficiently using 
available resources, being scalable with the number of connected devices and 
allowing flexible resource allocation among the different user classes.
 \item MOMA can be easily integrated into LTE systems as it can be implemented
on a sub-band of the LTE resource grid without affecting the legacy connections
occupying the rest of the bandwidth. Furthermore, MOMA signals can be
transmitted using modulation and coding schemes (MCSs) and transport block (TB)
sizes that are taken from the LTE standard. Finally, MOMA can make use of some
advanced features of LTE such as transmission time interval (TTI) bundling.
 \item A narrow-band implementation of MOMA is possible, thus making it
compatible with low-cost battery constrained devices.
 \item MOMA exploits massive MIMO both to increase user multiplexing
capabilities and to simplify the
receiver structure. Thanks to these properties, low-complexity detection is used
for most of the users in MOMA while the more complex detection schemes are only
applied for the highest data rate classes.
 \item Using MOMA entails a gain in coverage, making it advantageous for
connecting devices deployed in remote or bad-coverage areas.
 \item Several random access mechanisms with different degrees of signaling
overhead are compatible with MOMA.
\end{enumerate}
With these features, which are discussed in more detail in the sequel, MOMA can
overcome the shortcomings of the current proposals for the evolution of LTE in
supporting IoT. For instance, features 1, 4 and 6 are essential for enabling
real massive M2M deployments.

\subsection*{Service Dependent Hierarchical Spreading}

For the sake of clarity, we only consider the case $L=3$ from now on. The first
LMD class ($l=1$) will be simply referred to as the LD class, where LD stands
for {\em low data rates}, with target data rate 
$r^{\mathrm{LD}}$. Similarly, the second LMD class ($l=2$) will be referred to
as the MD class, where MD stands for {\em moderate data rates}, with target
data rate $r^{\mathrm{MD}}$.

To get a more precise description of MOMA, let $\mathbf{U}$ be a $N\times N$
code matrix (e.g. a Walsh-Hadamard matrix or a discrete Fourier transform
matrix). In MOMA, the set of columns of matrix $\mathbf{U}$ is divided into two
disjoint subsets, namely matrices $\mathbf{U}^{\mathrm{MD}}$ and
$\mathbf{U}^{\mathrm{LD}}$ with dimensions $N\times N^{\mathrm{MD}}$ and
$N\times N^{\mathrm{LD}}$ respectively. Now assume that a maximum number
$K^{\mathrm{MD}}$ ($K^{\mathrm{LD}}$) of simultaneously connected MD (LD) users
are to be served within the current sub-frame. Since we want to overload the MD
and LD radio resources, we will typically have $K^{\mathrm{MD}}>N^{\mathrm{MD}}$
and $K^{\mathrm{LD}}>N^{\mathrm{LD}}$. Finally, since we want to guarantee
higher data rates for the MD class as compared to the LD class, we impose 
$K^{\mathrm{LD}}/N^{\mathrm{LD}}>K^{\mathrm{MD}}/N^{\mathrm{MD}}$.

Instead of assigning the orthogonal spreading codes to individual users,
the $N^{\mathrm{MD}}$ (respectively $N^{\mathrm{LD}}$) columns of
$\mathbf{U}^{\mathrm{MD}}$ (respectively $\mathbf{U}^{\mathrm{LD}}$) are
simultaneously used in MOMA by the $K^{\mathrm{MD}}$ (respectively the
$K^{\mathrm{LD}}$) users. Indeed, each MOMA
transmitter applies as spreading code a linear combination of the columns of the
code matrix corresponding to its class. The coefficients of this linear
combination serve as a signature sequence to separate the signals of the users
belonging to the same class. More precisely, the data symbols of each MD (LD)
user are spread using one column of the product matrix
$\mathbf{U}^{\mathrm{MD}}\mathbf{W}^{\mathrm{MD}}$
($\mathbf{U}^{\mathrm{LD}}\mathbf{W}^{\mathrm{LD}}$) where
$\mathbf{W}^{\mathrm{MD}}$ ($\mathbf{W}^{\mathrm{LD}}$) is an
{\it overloading} matrix of dimensions $N^{\mathrm{LD}}\times K^{\mathrm{LD}}$
(respectively $N^{\mathrm{LD}}\times K^{\mathrm{LD}}$) whose columns are
referred to in the sequel as the \emph{overloading sequences}. In principle,
$\mathbf{W}^{\mathrm{MD}}$ ($\mathbf{W}^{\mathrm{LD}}$) can be
constructed by selecting $K^{\mathrm{MD}}$ ($K^{\mathrm{LD}}$)
points from the surface of the $N^{\mathrm{MD}}$-dimensional
($N^{\mathrm{LD}}$-dimensional) complex sphere with radius 1.
The resulting spread symbols of each user are then mapped to the elements of the
OFDMA/SC-FDMA time-frequency grid elements that fall within the frequency band
assigned to the MD and LD classes before being transmitted on the radio channel.
Finally, since the BS is equipped with a number $M>1$ of antennas, we know from
the literature~\cite{mimo_cdma_tse} that the effective spreading gain of MD
transmissions (respectively LD transmissions) is, roughly speaking,
proportional to $M N^{\mathrm{MD}}$ (respectively $M N^{\mathrm{LD}}$).
This intuition was confirmed by the analysis done in~\cite{eucnc_2016}.

The main advantage of this multiple-access scheme is an \emph{efficient,
scalable and flexible}  use of the available radio resources.
MOMA {\it efficiency} is shown by its overloading of the available
radio resources in order to connect a large number of IoT machines and of
handhelds requiring low-to-moderate data rates.
The {\it scalability} of MOMA with respect to increasing device densities is
simply a matter of applying a larger value for the MD (respectively LD)
overloading factor defined as $K^{\mathrm{MD}}/N^{\mathrm{MD}}$
(respectively $K^{\mathrm{LD}}/N^{\mathrm{LD}}$) and/or of employing a larger
$N$. MOMA {\it flexibility} is manifested by the ease
with which the network can dynamically adjust the proportion of resources
assigned to each class of devices/services and the degree to which these
resources are overloaded within each class by means of simply updating the
values of parameters $N^{\mathrm{MD}}$, $K^{\mathrm{MD}}$, $N^{\mathrm{LD}}$ and
$K^{\mathrm{LD}}$.
Finally, by properly mapping MOMA signals to the LTE time-frequency resource
grid, MOMA combines the benefits of both OFDM, e.g. robustness
against timing errors, and frequency-domain spreading, e.g. the
ability to harvest the frequency diversity of the channel and the robustness
against carrier frequency shifts.


\section*{MOMA Integration into Future LTE}

Let $B$ be the total system bandwidth and denote by
$B^{\mathrm{HD}}$ ($B^{\mathrm{LMD}}$) the bandwidth assigned to
the HD (MD and LD) class such that $B^{\mathrm{HD}}+B^{\mathrm{LMD}}=B$.
In order to apply MOMA in the future evolution of LTE, we need to set
  $B^{\mathrm{HD}}$ and $B^{\mathrm{LMD}}$, the spreading factor
$N$ and the map of the spread data symbols to the OFDMA/SC-FDMA
time-frequency grid. We also need to determine which new signaling messages
could be needed and what effect MOMA could have on the LTE system protocols.
First, let us recall how the available time-frequency resources are structured
into sub-frames in LTE. The smallest item in the time-frequency
grid in LTE is the resource element (RE) defined as one subcarrier within one
OFDM symbol of a duration equal to 66.7 $\mu$s. However, the basic unit for
scheduling and resource allocation is the physical resource block (PRB), which
is composed of $12$ REs in 14 consecutive OFDM symbols covering 180 kHz over
1 ms. The duration of the basic period of data scheduling in LTE,
called a sub-frame, is also equal to 1 ms.
Finally, the duration of one sub-frame is also
referred to as the data transmission time interval (TTI).

\subsection*{MOMA on a 1.4 MHz Bandwidth}

\begin{figure}[h]
 \centering
 \includegraphics[width=0.8\hsize]{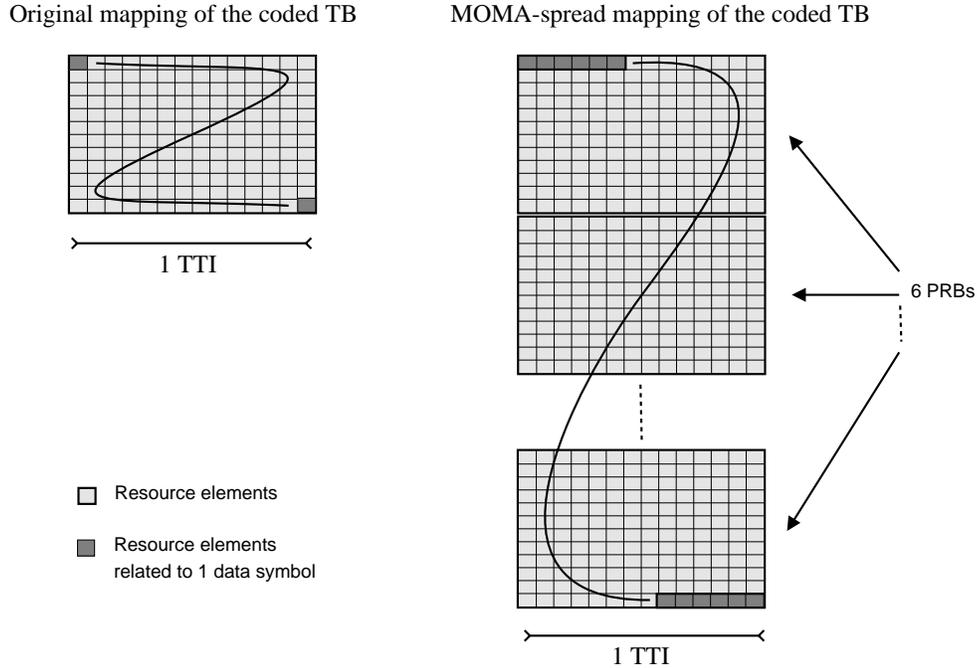}
 \caption{Implementing MOMA on 6 PRBs.}
 \label{fig:tb_bundling_off}
\end{figure}

One possible implementation of MOMA provides that $B^{\mathrm{LMD}}$ coincides
with the frequency band occupied by the 6 PRBs that are destined for M2M
communications in LTE-M. This implementation is illustrated in
Fig.~\ref{fig:tb_bundling_off}. 
In this implementation, at the beginning of each TTI each active MD and LD
transmitter extracts from its transmission queue enough data bits that, after
coding and mapping, results in a number of data symbols equal to the size of
one PRB. The motivation behind this choice is to retain full compatibility with
LTE MCSs and TB sizes. Next, each data symbol is spread using
service dependent hierarchical spreading with $N=6$. The resulting spread symbol
is finally mapped to 6 consecutive REs from one PRB as shown in
Fig.~\ref{fig:tb_bundling_off}.

\subsection*{MOMA on a 200 kHz Bandwidth}

Another possible MOMA implementation is obtained by letting
$B^{\mathrm{LMD}}$ coincide with the one PRB reserved for M2M
communications in NB LTE-M. Except for the difference in bandwidth and in
resource allocation granularity, this implementation is not different from the
1.4 MHZ implementation.

\subsection*{MOMA with TTI Bundling}

\begin{figure}[h]
 \centering
 \includegraphics[width=0.8\hsize]{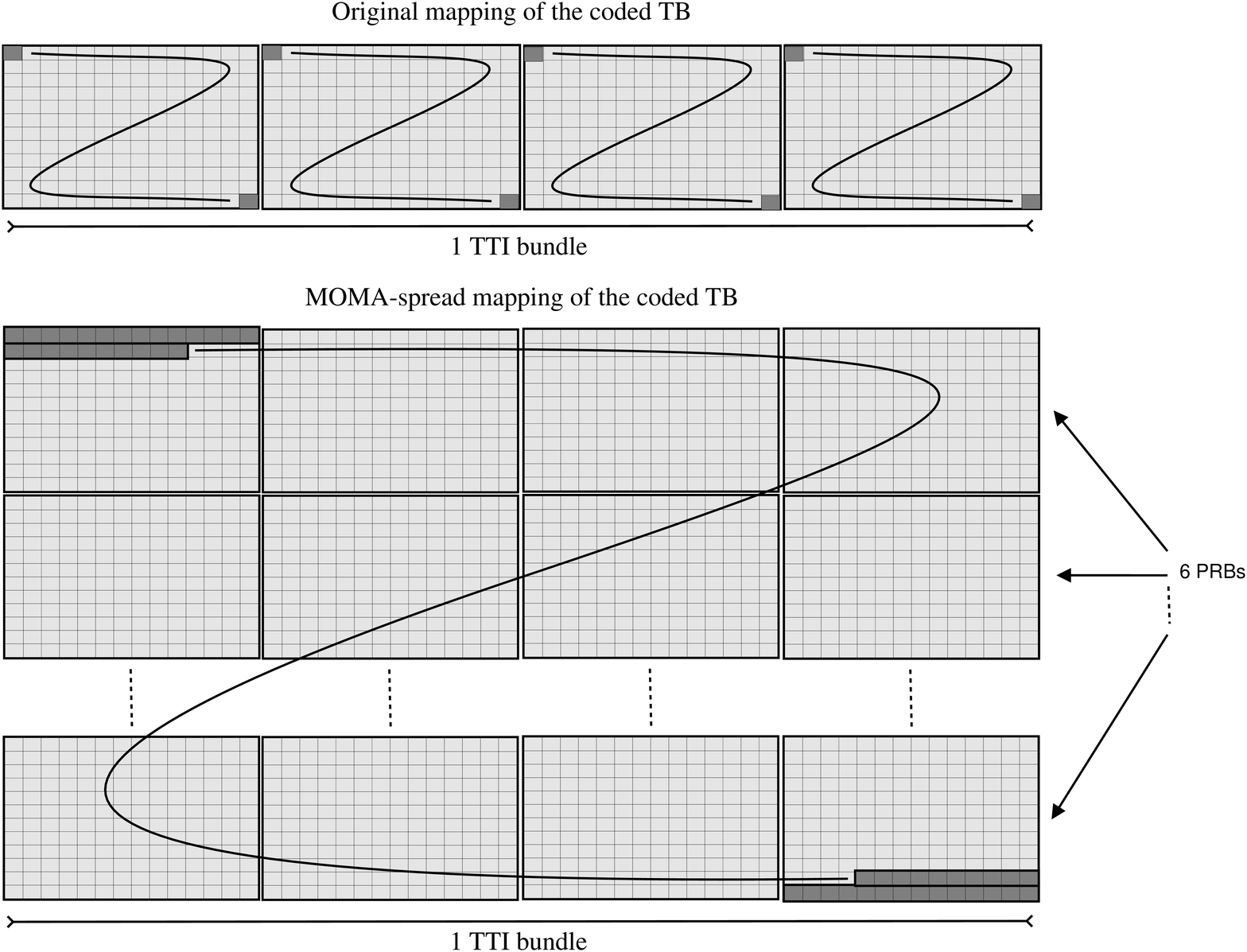}
 \caption{Implementing MOMA on 6 PRBs with TTI bundling.}
 \label{fig:tb_bundling_on}
\end{figure}

TTI bundling is a transmission technique that was originally proposed for
coverage enhancement in delay-limited applications such as Voice-over-LTE
(VoLTE)~\cite{coverage_enhancement}. It consists in allowing users to
transmit the four redundancy versions (RVs) of their current codewords in one
shot using four consecutive TTIs, instead of waiting for the BS
acknowledgment (ACK) or negative acknowledgment (NACK) message after each RV
transmission. When applied along with MOMA, the already-existing control
messages and protocol structure that were introduced to support TTI bundling can
be re-purposed in order to increase the multiplexing capabilities for IoT
devices while getting a coverage enhancement gain.
In a 1.4 MHz implementation of MOMA with TTI bundling, the MD and LD spreading
code length is $4N$, where $N$ is the original spreading factor without TTI
bundling. Each sequence resulting from spreading a MD or a LD symbol is mapped
to $4N$ consecutive REs from one PRB as shown in Fig.~\ref{fig:tb_bundling_on}.

\subsection*{MOMA Receiver Complexity and Performance} 

We propose a receiver structure for MOMA that takes advantage of the large
number $M>>1$ of BS antennas. This massive-MIMO scenario, which is expected to
be prevalent in next-generation cellular networks, proves to be advantageous for
MOMA from both performance and receiver complexity perspectives.
The proposed receiver is illustrated in the right-hand part of
Fig.~\ref{fig:moma_principle} and consists in performing the following
two steps.

{\bf Spatial combining:}
When the BS has multiple antennas, linear receive combining in the uplink is
typically utilized. When the number of BS antennas $M$ is large enough,
maximum-ratio combining (MRC), which is a low-complexity scheme, has been shown
\cite{mimo_myths} to achieve a spectral efficiency not far from that achieved
with more involved linear combining methods thanks to the asymptotic (with
respect to the number of BS antennas) orthogonality of users' channel vectors in
massive MIMO. We thus propose to apply MRC for the detection of both LD and MD
spread signals on the {\em chip level}.

{\bf Code despreading:}
Since the target data rate $r^{\mathrm{MD}}$ for the MD class is
relatively high as compared to $r^{\mathrm{LD}}$, the number of devices that can
be simultaneously served in that class is expected to be smaller than its LD
counterpart. The BS can thus typically afford for this class the use of
multiuser detection techniques such as successive interference cancellation
(SIC). On the other hand, we will consider only single-user detection for the LD
class. This choice is motivated by the need to maintain a reasonable detection
complexity while serving a large number of LD users with low target data rates.

\begin{figure}[h]
 \centering
 \includegraphics[width=0.8\hsize]{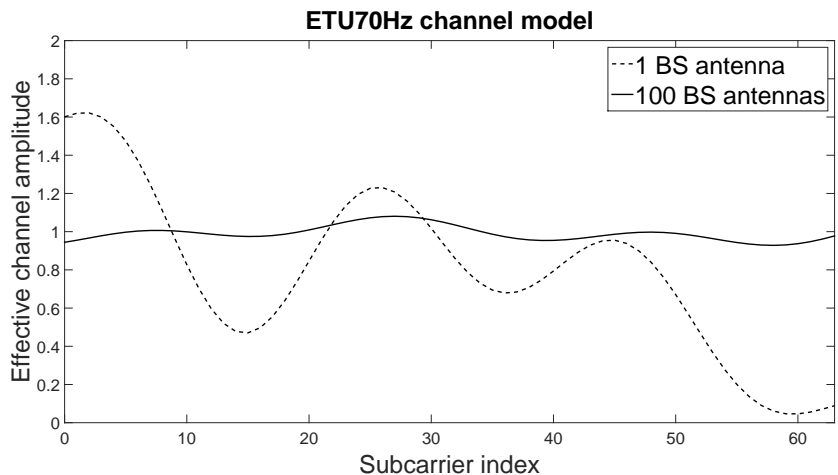}
 \caption{Channel hardening effect in MRC combining.}
 \label{fig:channel_hardening}
\end{figure}

Interestingly, this simple receiver structure which does not involve any
inter-class multiuser detection was shown~\cite{eucnc_2016} to achieve
asymptotic MD/LD orthogonality even on fast-varying
frequency-selective channels and even when the number $K^{\mathrm{LD}}$ of LD
users grows to infinity. This is due to a favorable property of massive-MIMO
channels. Indeed, as an effect of combining a large number $M$ of
signals in the case of a rich-scattering propagation environment,
the small-scale fading averages out over the array in the sense that the
variance of the resulting scalar channel decreases with $M$. This effect is
known as {\it channel hardening} and is a consequence of the law of large
numbers~\cite{mimo_myths}. Most importantly in our case, the frequency response
of the effective channel is asymptotically flat and asymptotically constant over
several consecutive OFDM symbols as illustrated in
Fig.~\ref{fig:channel_hardening} when  $M=100$. The channel
realizations used in this figure were generated using the Extended Type Urban
(ETU) channel model~\cite{3gpp_etu_epa}.

\begin{figure}[h]
 \centering
 \includegraphics[width=1\hsize]{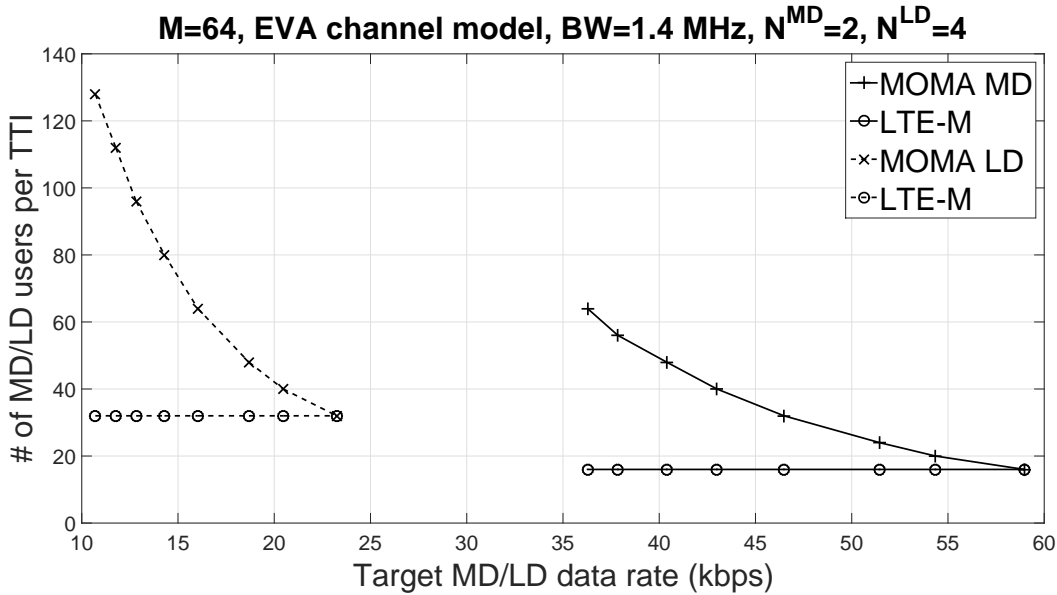}
 \caption{Number of served MD/LD users vs. target data rate. Without TTI
bundling.}
 \label{fig:kLD_rLD_no_bundling}
\end{figure}

\begin{figure}[h]
 \centering
 \includegraphics[width=1\hsize]{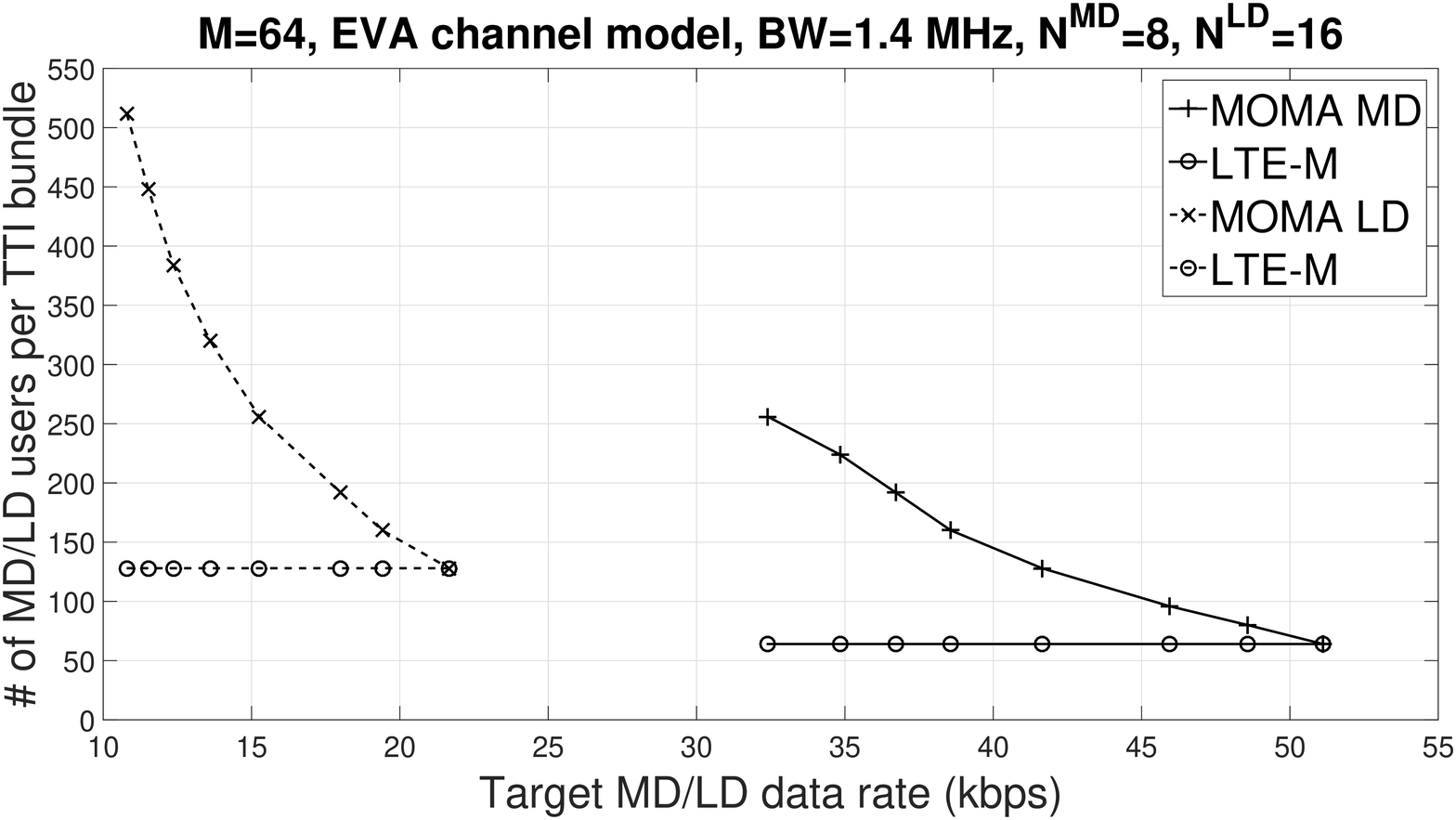}
 \caption{Number of served MD/LD users vs. target data rate. With TTI bundling.}
 \label{fig:kLD_rLD_bundling}
\end{figure}

In Fig.s \ref{fig:kLD_rLD_no_bundling} and \ref{fig:kLD_rLD_bundling}, we
plotted the number of MD and LD connections that can be simultaneously served
with and without TTI bundling, respectively, as function of their respective
target data rates ($r^{\mathrm{MD}}$ and $r^{\mathrm{LD}}$) for both MOMA and
LTE-M. The values of $r^{\mathrm{MD}}$ are taken from the range
$[30,60]$ kbps while $r^{\mathrm{LD}}\in[10,25]$ kbps. The higher value in
these two intervals is dictated by the maximum per-link data rate achievable
with orthogonal (thus underloaded) access schemes.
From the figures we can notice the significant advantage of using MOMA as
opposed to orthogonal-access NB CIoT solutions in terms of the capability of
serving densely-deployed IoT devices. For instance, four times more simultaneous
MD and LD connections can be served while both $r^{\mathrm{MD}}$ and
$r^{\mathrm{LD}}$ are approximately at half their respective upper-bound values.
Note that this performance has been obtained on doubly dispersive channels
generated using the Extended Vehicular A (EVA) model~\cite{3gpp_etu_epa} which
is characterized with a relatively long delay spread and short coherence time.
Also note that the four times larger spreading gain resulting from TTI bundling
allows to serve four times more MD and LD simultaneous connections  while
meeting their respective target data rates, in the same range as in the absence
of TTI bundling.
The figures were obtained using 100 realizations of users' distances to the BS
randomly chosen in the interval $\left[25,100\right]$~m assuming
that the BS is equipped with $M=64$ antennas and that users' transmit power is
equal to 23 dBm.

\subsection*{Coverage Enhancement with MOMA}

The coverage measure adopted for LTE channels is the {\em
maximum coupling loss} (MCL), defined as the difference in logarithmic
scale between the maximum transmission power and the receiver
sensitivity~\cite{coverage_enhancement}. A higher MCL value indicates that
the transmitter-receiver distance can be made larger while still meeting
the target received signal-to-noise ratio (SNR) (and hence the target block
error rate (BLER)). This translates into better cellular service coverage.

As a consequence of the $N$-long spreading used in MOMA, IoT users benefit
from a coverage enhancement gain. This gain can be further increased if
TTI-bundling is applied due to the higher processing gain (equal to $4N$)
resulting in this case from spreading over a larger number of resource blocks.
Indeed the increase in MCL due to the use of MOMA is in the order of
$10\log_{10} N$ without TTI bundling and of $10\log_{10}(4N)$ with TTI bundling,
corresponding to a gain of 7.78 dB and 13.80 dB, respectively.

\subsection*{Signaling and Random Access for MOMA}

To implement MOMA, matrix $\mathbf{U}$ should be known in advance to all 
MD and LD users, for instance in the form of a look-up table (LUT).
Moreover, the resource allocation parameters $N^{\mathrm{MD}}$ and
$N^{\mathrm{LD}}$ are dynamic and need to be broadcast by the BS.
This can be done by making use of the LTE broadcast control channel (BCH).
Note that the values of $N^{\mathrm{MD}}$ and $N^{\mathrm{LD}}$ can typically be
kept fixed for a relatively long time as they only need to
be changed when the traffic characteristics change. As for the overloading
sequences, their allocation is intimately related to the kind of random
access (RA) scheme we opt for.

{\bf Contention-free RA:} In a contention-free access scheme, the BS is in full
control of the assignment of the available radio resources to the active users
in the cell area~\cite{lte_ra}. In MOMA, this translates into the BS choosing
which MD (respectively LD) user is assigned which column of the overloading
matrix $\mathbf{W}^{\mathrm{MD}}$ (respectively $\mathbf{W}^{\mathrm{LD}}$).
Note that these two matrices could be generated in advance, i.e. offline, for
different combinations of the values of $N^{\mathrm{MD}}$, $N^{\mathrm{LD}}$,
$K^{\mathrm{MD}}$ and $K^{\mathrm{LD}}$ and made available in the form of a
LUT to the relevant users. Contention-free RA has the advantage of eliminating
collisions among concurrent connections, and hence eliminating the need for
collision detection and resolution. However, this comes at the price of a
relatively large protocol overhead, especially in the case of a highly
overloaded system. Contention-free RA is thus more suitable for systems with
low-to-moderate user densities. 

{\bf Contention-based RA:} In a contention-based access scheme, we let MD
and LD users compete within their respective classes for the overloading
sequences. On one side this helps cut the protocol overhead thus making
contention-based RA relevant for systems with high user densities. On the other
side, it comes at the price of eventual collisions between concurrent uplink
transmissions and ensuing retransmissions. A conflict/collision occurs when at
least two MD or LD users choose the same overloading sequence, making it
difficult for the BS to correctly detect their transmitted data symbols.
In contention-based MOMA, there is no need to store LUTs corresponding to the
overloading matrices as the overloading sequences could be locally generated
when needed. In this case, collision resolution is left entirely to the
higher network layers. Otherwise, the probability of collisions could be reduced
by using preamble transmission as in LTE PRACH and/or by using contention
transmission unit (CTU) messages \cite{scma_ra} that have the overloading
sequence as one of their fields.

{\bf Hybrid RA scheme}: Interestingly, LD collisions that take place in
contention-based MOMA cannot affect MD connections and both MD and LD collisions
cannot affect HD (legacy) connections thanks to inter-class quasi-orthogonality
in MOMA. This observation can be used to motivate the use of a contention-free
RA scheme for HD and MD connections and a contention-based RA scheme for LD
connections. Such a hybrid scheme has the advantage of reducing protocol
overhead in systems characterized with a relatively high LD and a lower MD
device density.

Note that in all three cases uplink synchronization is needed before
initiating the RA procedure and the actual data transmission. As in LTE,
this can be maintained with a timing advance procedure~\cite{lte_book}.

\section*{Conclusions and Perspectives}

MOMA is a novel multiple access scheme compatible with massive MIMO, 
which can be integrated into the evolution of LTE to enhance its support for a
wide range of services including M2M communications. MOMA is based on assigning,
in a flexible and dynamic manner, different code resources and different degrees
of resource overloading to different classes of users, each representing a
different data rate requirement, a different service type and/or a different
traffic pattern. Code assignment in MOMA is conceived in such a way that
overloading the resources of the lower data rate classes would only slightly
affect the higher data rate classes, dropping the need for wasteful guard bands
and steep transmit filters for uplink transmission. Moreover, the different QoS
requirements in the system can be satisfied in a flexible and efficient fashion
by reserving higher-complexity detection schemes at the BS only for classes
which need them. Finally, we showed that MOMA outperforms the other M2M-related
proposals for LTE in most of the M2M-relevant performance measures while
requiring comparable signaling and protocol overhead.

One research direction for MOMA consists in analyzing its performance using
different MIMO channel models reflecting the diverse propagation
environments and practical BS array configurations.
Another research direction is conducting a higher-level assessment of MOMA
using advanced models, as those introduced in~\cite{traffic}, for the data
traffic generated by the different classes of users. Finally, the integration
into MOMA of other coverage enhancement techniques such as relaying is yet to be
investigated.


\begin{IEEEbiographynophoto}{Nassar Ksairi}
received the M.Sc. degree from CentraleSupelec, France, in 2006 and the Ph.D.
degree from the University of Paris-Sud XI, France, in 2010. From 2010 to 2012,
he was an Assistant Professor at the Higher Institute for Applied Sciences and
Technology, Damascus, Syria. From 2012 to 2014, he was a Postdoctoral researcher
at T\'el\'ecom ParisTech, Paris, France. Since December 2014, he is a researcher
at Huawei's Mathematical and Algorithmic Sciences Lab, France.
\end{IEEEbiographynophoto}

\begin{IEEEbiographynophoto}{Stefano Tomasin}
received his PhD in 2003 from the University of Padova, Italy, which he then
joined since 2002, being now Associate Professor. He has spent leaves and
sabbaticals at Qualcomm in California, the Polytechnic University in New York
and the Mathematical and Algorithmic Sciences Laboratory of Huawei Technologies,
in France. Since 2011 he is Editor of both IEEE Transactions of Vehicular
Technologies and EURASIP Journal of Wireless Communications and Networking.
\end{IEEEbiographynophoto}

\begin{IEEEbiographynophoto}{M\'erouane Debbah}
entered the Ecole Normale Sup\'erieure de Cachan (France) in 1996 where he
received his M.Sc and Ph.D. degrees respectively. He worked for Motorola Labs
(Saclay, France) from 1999-2002 and the Vienna Research Center for
Telecommunications (Vienna, Austria) until 2003. From 2003 to 2007, he joined
the Mobile Communications department of the Institut Eurecom (Sophia Antipolis,
France) as an Assistant Professor. Since 2007, he is a Full Professor at
CentraleSupelec (Gif-sur-Yvette, France). From 2007 to 2014, he was the director
of the Alcatel-Lucent Chair on Flexible Radio. Since 2014, he is Vice-President
of the Huawei France R\&D center and director of the Mathematical and
Algorithmic Sciences Lab.
\end{IEEEbiographynophoto}


\end{document}